\documentclass[12pt]{article}
\input epsf.sty
\def\be{\begin{equation}}
\def\ee{\end{equation}}
\def\ba{\begin{align}}
\def\ea{\end{align}}

\def\p{\partial}

\def\ops[#1]{\p_{#1} e^{-2\phi}}

\def\eq[#1]{equation (\ref {eq:#1})}
\def\Eq[#1]{Equation (\ref {eq:#1})}
\def\e[#1]{\ref {eq:#1}}
\def\at[#1]{| _{#1}}

\let\oldpercent\%\renewcommand{\%}{\scalebox{0.85}{\oldpercent}}

\begin{document}

\baselineskip=18pt

\begin{center}
{\Large \bf{On string theory on deformed BTZ and $T\bar T$}}

\vspace{10mm}

\textit{Soumangsu Chakraborty$^{a}$, Amit Giveon$^b$, Akikazu Hashimoto$^c$}
\break

$^a$ Institut de Physique Th\'eorique, Universit\'e Paris-Saclay, CNRS, CEA\\
Orme des Merisiers, 91191 Gif-sur-Yvette, France\\
$^b$Racah Institute of Physics, The Hebrew University \\
Jerusalem, 91904, Israel\\
$^c$ Department of Physics, University of Wisconsin\\
Madison, WI 53706, USA

\vspace{6mm}

$^a$soumangsuchakraborty@gmail.com, \\$^b$giveon@mail.huji.ac.il, \\$^c$aki@physics.wisc.edu

\end{center}

%\break

\vspace{3mm}

\begin{abstract}

Aspects of superstring theory on deformed BTZ black holes, formed near $k$~NS5 branes by $p$~fundamental strings,
and single-trace $T\bar T$ holography, are presented.
The excitation energy of a singly wound long string plus its contribution to the energy of the black hole,
$1/p$ fraction of the black-hole energy, is the same as that in $T\bar T$ deformed $CFT_2$ with $c=6k$.
This supports the claim that the black hole can be thought of as a state in a symmetric product of $p$~$CFT_2$'s
of central charge~$6k$, where  the black-hole energy is split equally among all $p$ factors.
A comment on superstring theory on deformed global $AdS_3$ is also presented.

\end{abstract}
\vspace{10mm}

\section{Introduction}

String theory on $AdS_3 \times {\cal N}$ is an important conceptual laboratory (see e.g.~\cite{Aharony:1999ti} for a review).
If constructed using $k$ NS5 branes and $p$ fundamental strings, the geometry is supported by NS-NS $B$-field.~\footnote{Though
we emphasize the $k$ NS5 branes on $T^4$ example, in which case ${\cal N}=S^3\times T^4$,
our results below apply to perturbative superstring theory on a level $k$ $AdS_3$ for any internal space ${\cal N}$,
e.g. the ones in~\cite{Giveon:1999jg}.}
The worldsheet sigma-model for the $AdS_3$ sector can be constructed explicitly as a WZW model.
The spacetime theory is holographically interpretable as a two-dimensional conformal field theory.
If the theory is specified on the string theory side, the holographically dual $CFT_2$ is also uniquely specified.
It is generally believed that this theory shares many features with a $p$-fold symmetric product~\cite{Argurio:2000tb}.
Specifically, the spectrum of long strings appears to exhibit a symmetric product structure ${\cal M}^p/S_p$,
where the block $\mathcal{M}$ is a $CFT_2$ of central charge~$6k$,
describing the target-space dynamics of a single long string~\cite{Giveon:2005mi}.
However, this fact alone does not imply that the full boundary CFT is a symmetric product theory,
since there are additional discrete states which do not fall into this category (see e.g.~\cite{Chakraborty:2023mzc}
for the present status of the above).

One approach to probe the physical content of $AdS_3$ holography is to explore a larger set of theories connected by a deformation parameter. One such family was constructed in~\cite{Giveon:2017nie,Giveon:2017myj}.
The deformation considered in~\cite{Giveon:2017nie,Giveon:2017myj} is an exactly marginal deformation from the worldsheet point of view. From the $2d$ holographic CFT point of view, the deformation is irrelevant. In the bulk description, the background geometry has an interpolation of locally $AdS_3$ in the core region smoothly connected to flat space-time, $R_t\times S^1\times R_\phi$ with a linear dilaton.
The curvature of the $AdS_3$ cap, $1/\sqrt{\alpha'k}$, is also the background charge of the linear dilaton,
and we denote the radius of the asymptotic $S^1$ by $R$.

The UV dynamics of the deformed CFT exhibit non-locality of little string theory~\cite{Aharony:1998ub}.
Furthermore, the target-space dynamics of a single probe long string in the defomed geometry exhibit a $T \bar T$  deformation (see e.g.~\cite{Giveon:2017myj,Hashimoto:2019wct}).
For $N$ such free long strings, the target-space dynamics is that of a $N$-fold symmetric product of $T\bar{T}$ deformed single long string CFT.
Consequently, it suggested that in the long strings sector,
the deformation of~\cite{Giveon:2017nie,Giveon:2017myj} should be understood as the $p$-fold symmetric product
of a $T \bar T$ deformed CFT seed, ${\cal M}$, with $c=6k$.
This duality is referred to as ``single-trace $T\bar T$ holography.''~\footnote{See e.g.~\cite{Chakraborty:2023mzc}
for its present status.}

Another interesting probe of string theory on $AdS_3 \times {\cal N}$
and its single-trace $T\bar T$ deformation is to explore the thermodynamic properties at finite temperature, e.g. for black-hole states.
In~\cite{Giveon:2017nie} (see also \cite{Chakraborty:2023mzc,Chakraborty:2023zdd} and references therein),
it was found that the deformation of~\cite{Giveon:2017nie,Giveon:2017myj} acted in such a way that the $1/p$ fraction
of the dimensionless black-hole energy,
\be\label{ebhp} \frac{E_{BH}R}{p}, \ee
gets modified  by the deformation as if $1/p$ fraction of the total dimensionless energy of the undeformed BTZ black hole with mass $M_{BTZ}$,
\be\label{mbtzp} \frac{\sqrt{\alpha' k}M_{BTZ}}{p}, \ee
is transforming according to the universal $T \bar T$ deformation formula~\cite{Smirnov:2016lqw,Cavaglia:2016oda}.
This suggests that a typical BTZ black-hole state can be thought of as a state in a $p$-fold symmetric product $CFT_2$ with total central charge~$6kp$, where the total energy is equally distributed among the $p$~copies, and that the deformation we are interested in acts as a $T\bar{T}$ deformation of the individual block CFT.

In this note, we will extend this line of thinking, and consider the spectrum of a perturbative string excitation in the black-hole background.
We will focus on long string states.

In computing the spectrum, it is important to know the value of $B_{\tau \theta}(\rho=0)$, at the origin of the BTZ black hole, where $\tau$, $\theta$ and $\rho$ are the time, angular and radial coordinates of the BTZ black hole, respectively (see section 2).
Although a constant shift of $B$ does not affect $H=dB$, and therefore the supergravity equations of motion, it does affect the energies of the strings in a way that depends on its winding number.
This issue was discussed recently from a different point of view in~\cite{Martinec:2023plo} (see also~\cite{Ashok:2021ffx}).
It was argued there that $B_{\tau\theta}(\rho=0)=0$.~\footnote{The relevant discussion can be found below (2.11) in \cite{Martinec:2023plo}.}  In this note, we will explore the spectrum that follows specifically from the prescription of \cite{Martinec:2023plo}.

Our main result thus obtained is the following (see sections 2 and 3).
The energy spectrum of wound strings in the deformed theory,
\begin{equation}\label{ewtotal}
E_{w,total}R \equiv E_wR + w \frac{E_{BH}R}{p},
 \end{equation}
with undeformed spectrum
\begin{equation}
   {\cal E}_{w,total} \equiv {\cal E}_w + w \frac{\sqrt{\alpha'k}M_{BTZ}}{p},
\end{equation}
satisfies the $T\bar T$ formula of~\cite{Smirnov:2016lqw,Cavaglia:2016oda}, for a theory on a circle with radius $R$,
applied to a $p$-fold symmetric product of a $T\bar T$ deformed $CFT_2$ with $c=6k$.
Here, $E_w$ is the energy of the probe long string with winding number $w$,
and $E_{BH}$ is the energy of the deformed BTZ black hole.
$E_{w,total}$ is the sum of the contribution of a long string excitation
and the $w/p$ fraction of the full background contribution to the energy.
The symbol ${\cal E}$ refers to the $R/\sqrt{\alpha'}\rightarrow\infty$ limit of the dimensionless energy, $ER$.

This result is in harmony with the single-trace $T \bar T$ holography conjecture
of~\cite{Giveon:2017nie,Chakraborty:2020swe,Chakraborty:2023mzc,Chakraborty:2023zdd}.
{\it{A key insight is the fact that considering a $w$-wound string state as an excitation
is accompanied by the back reaction to the background, shifting $p$ to $p-w$.}}
In the remainder of this article, we will provide the details of the analysis which lead to the conclusion summarized above.

\section{Spectrum of the long strings}

Consider supersting theory on the deformed BTZ black holes background,
formed in the near $k$ NS5 branes on $S^1\times T^4$
with $p$ fundamental strings wrapping $S^1$, whose radius we denote by $R$.~\footnote{For simplicity,
we consider the case without fundamental strings momentum on $S^1$, which gives rise to non-rotating black holes, $J=0$.}
This background appears in e.g. equations (2.16) with (2.17) of~\cite{Chakraborty:2023mzc}:
\be\label{defbtz}
ds^2=-{N^2\over 1+{\rho^2\over R^2}}d\tau^2+{\rho^2\over 1+{\rho^2\over R^2}}d\theta^2+{d\rho^2\over N^2},\qquad \theta\simeq\theta+2\pi,
\ee
\be\label{btautheta}
B_{\tau\theta}={\rho^2\over r_5}\sqrt{1+{\rho_0^2\over R^2}}{1\over 1+{\rho^2\over R^2}},
\ee
\be\label{dilaton}
e^{2\Phi}={kv\over p}\sqrt{1+{\rho_0^2\over R^2}}{1\over 1+{\rho^2\over R^2}},\qquad v\equiv {\rm Volume}(T^4)/\left(2\pi\sqrt{\alpha'}\right)^4,
\ee
with~\footnote{In~\cite{Chakraborty:2023mzc},
the angular direction $\theta$ of the background~(\ref{defbtz}),(\ref{btautheta}) was denoted by $\varphi$;
below, $\varphi,\bar\varphi$ will denote worldsheet fields associated with the superconformal ghosts, instead.}~\footnote{The
integration constant in going from $H$ to $B$ is not ambiguous; as in~\cite{Chakraborty:2023mzc,Chakraborty:2023zdd}, it is fixed
such that the $B$-field in~(\ref{btautheta}) vanishes at $\rho=0$.
This is compatible with the choice argued for in~\cite{Ashok:2021ffx,Martinec:2023plo}.}
\be\label{withn}
N^2={\rho^2-\rho_0^2\over r_5^2},
\ee
and
\be\label{k}
r_5\equiv\sqrt{\alpha' k}.
\ee
At $\rho/R\to\infty$, the background~(\ref{defbtz})--(\ref{dilaton}) with~(\ref{withn})  takes the form
\be\label{ds}
ds^2\to -dt^2+dx^2+d\phi^2,\qquad B_{tx}\to b\equiv\sqrt{1+{\rho_0^2\over R^2}},\qquad\Phi\to -{\phi\over r_5}+const,
\ee
where
\be\label{tx}
t\equiv{R\tau\over r_5},\qquad x\equiv R\theta,\qquad{\phi\over r_5}\equiv\log\left(\rho\over R\right).
\ee
In other words, it is time times a circle, $R_t\times S^1_{x\simeq x+2\pi R}$, with a constant $B$-field,
times an $R_\phi$ with a linear dilaton.

A large class of observables in the (NS,NS) sector of superstring theory on deformed BTZ is given by vertex operators
in the $(-1,-1)$ picture whose behavior in the asymptotically linear dilaton regime,~(\ref{ds}), is
\be\label{v}
V_{phys}\to e^{-\varphi-\bar\varphi}V_{N_L,N_R}e^{-iE t}e^{ip_Lx_L+ip_Rx_R}e^{2j\phi/r_5},
\ee
where
\be\label{plpr}
(p_L,p_R)=\left({wR\over\alpha'}+{n\over R},{wR\over\alpha'}-{n\over R}\right),
\ee
and
\be\label{onshell}
{\alpha'\over 4}\left(-(E+bwR/\alpha')^2+p_{L,R}^2\right)-{j(j+1)\over k}+N_{L,R}={1\over 2}.
\ee
Each of these operators amounts to a string with winding $w$ and momentum $n$ on the asymptotic circle, $S^1_{x\simeq x+2\pi R}$,
moving with momentum governed by the quantum number $j$ in the radial direction,
in a particular state of transverse left and right-handed levels, $N_{L,R}$;
eq.~(\ref{onshell}) is obtained from the mass-shell relation of physical vertex operators
and using e.g. (2.4.11) in~\cite{Giveon:1994fu}~\footnote{More details:
for canonically normalized time, $t$,
times a circle, parameterized by $y\sim y+2\pi$,
with radius $r_y$ in $\sqrt{\alpha'}$ units,
manipulations in~\cite{Giveon:1994fu}, adapted to time times a circle, give
$$
\Delta+\bar\Delta={1\over 2}\left(n_i(g^{-1})^{ij}n_j+m^i(g-bg^{-1}b)_{ij}m^j+2m^ib_{ik}(g^{-1})^{kj}n_j\right),
$$
with
$$
g_{ij}=\pmatrix{
 -1&0\cr
 0&r_y^2&
},\qquad
b_{ij}=\pmatrix{
0 & -br_y\cr
br_y& 0
},
$$
$$
n_i=(e,n_y),\qquad m^i=(0,w_y),\qquad i,j=t,y,
$$
for (dimensionless) energy $e\equiv\sqrt{\alpha'}E$ and momentum and winding $n_y,w_y$ on $S^1_y$,
from which one finds
$$
\Delta+\bar\Delta={1\over 2}\left(-(e+br_yw_y)^2+{n_y^2\over r_y^2}+w_y^2r_y^2\right)~,\qquad \Delta-\bar\Delta=n_yw_y,
$$
which matches the first term in~(\ref{onshell}).
}
(adapted to~(\ref{ds})).

Equation~(\ref{onshell}), with the $b$ in~(\ref{ds}), can be rewritten as~\footnote{The label $w$ is added to the $E$ above for convenience later.}
\be\label{e}
E_w+{wR\over\alpha'}\left(-1+\sqrt{1+{\rho_0^2\over R^2}}\right)=
-{wR\over\alpha'}+\sqrt{{w^2R^2\over\alpha'^2}+{2\over\alpha'}\left(-{2j(j+1)\over k}+N_L+N_R-1\right)+{n^2\over R^2}}.
\ee
For long strings with radial momentum ${2s\over r_5}$, namely, with $j=-{1\over 2}+is$, $s\in R$,
we can keep $j$ and all the other quantum numbers fixed, as we vary $R$, thus obtaining a flow of energies, $E_w(R)$,
for each winding number, $w=1,2,3,\dots$, in eq.~(\ref{e}).

The undeformed BTZ background and the long strings spectrum of the superstring on BTZ are obtained by considering the background~(\ref{defbtz})--(\ref{withn})
in the limit $\rho_0,\rho\ll R$,~\footnote{See eqs. (2.25)--(2.27) and the discussion around them in~\cite{Chakraborty:2023mzc}.}
and taking the $R/\sqrt{\alpha'}\to\infty$ limit of~(\ref{e}), respectively.
Now, following  e.g.~\cite{Martinec:2023plo} and references therein, for long strings in BTZ,~\footnote{See,
in particular, the discussion around eqs. (2.32)--(2.34) in~\cite{Martinec:2023plo}.}
\be\label{h}
-{2j(j+1)\over k}+N_L+N_R-1=w\left({\cal E}_{w,\rm string}+{w\rho_0^2\over 2\alpha'}\right),
\ee
where ${\cal E}_{w,\rm string}$ is the (dimensionless) perturbative excitation energy carried by a string carrying winding $w$
in the superstring on BTZ,~\footnote{See eq. (2.34) in~\cite{Martinec:2023plo}.}
and ${w\rho_0^2\over 2\alpha'}$ is the contribution to the energy of a BTZ black hole (in units of $1/r_5$)
due to $w$ fundamental strings on the $\theta$ direction of the BTZ limit of~(\ref{defbtz}),
out of the (parametrically large) $p$ fundamental strings that form it.~\footnote{See eq. (2.33) in~\cite{Martinec:2023plo}, and below.}
This is consistent with the $R/\sqrt{\alpha'}\to\infty$ limit of~(\ref{e}),
\be\label{de}
\lim_{{R\over\sqrt{\alpha'}}\to\infty}RE_w={\cal E}_{w,\rm string}.
\ee
For each set of $w=1,2,3,\dots$ long strings states with the same $j$ and other quantum numbers, we thus have, in particular,
\be\label{jj}
-{2j(j+1)\over k}+N_L+N_R-1={\cal E}_{\rm string}+{\rho_0^2\over 2\alpha'},
\ee
regardless of their winding number, $w$, where
\be\label{de1}
{\cal E}_{\rm string}\equiv{\cal E}_{1,\rm string}
\ee
is the energy carried by a string with $w=1$.

To recapitulate, we found that the long strings spectrum in superstring theory on deformed BTZ has the pattern:~\footnote{The
expression for probe string energy $E_w$ in \cite{Apolo:2019zai,Chakraborty:2020yka,Chakraborty:2022dgm} differ from~(\ref{recap}).}
%~\footnote{The results in \cite{Apolo:2019zai,Chakraborty:2020yka,Chakraborty:2022dgm} differ from~(\ref{recap}).}
\be\label{recap}
E_w(R)+{wR\over\alpha'}\left(-1+\sqrt{1+{\rho_0^2\over R^2}}\right)=
-{wR\over\alpha'}+\sqrt{{w^2R^2\over\alpha'^2}+{2\over\alpha'}
\left({\cal E}_{\rm string}+{\rho_0^2\over 2\alpha'}\right)
+{n^2\over R^2}},
\ee
where ${\cal E}_{\rm string}$ is the dimensionless energy carried by a winding one long string in the superstring on a BTZ black-hole CFT:
\be\label{estring}
{\cal E}_{\rm string}=\lim_{R\to\infty}RE_1(R).
\ee
Note that the addition to ${\cal E}_{\rm string}$, in the parenthesis inside the square root on the r.h.s. of~(\ref{recap}),
is the dimensionless BTZ black-hole mass (in $1/\sqrt{\alpha' k}$ units) divided by $p$:~\footnote{See
eqs. (2.27),(2.28) in~\cite{Chakraborty:2023mzc}.}
\be\label{mbtz}
{\rho_0^2\over 2\alpha'}={r_5M_{BTZ}\over p}.
\ee
Finally, note that the addition to $E_w(R)$ on the l.h.s. of~(\ref{recap})
is the energy of the deformed BTZ black hole divided by $p$:~\footnote{See eq. (2.20) in~\cite{Chakraborty:2023mzc}.}
\be\label{edefbtz}
{R\over\alpha'}\left(-1+\sqrt{1+{\rho_0^2\over R^2}}\right)={E_{\rm deformed\,BTZ}(R)\over p}.
\ee
Now, following the single-trace $T\bar T$ holographic dictionary  in~\cite{Giveon:2017nie,Giveon:2019fgr,Chakraborty:2019mdf,Chakraborty:2020swe,Chakraborty:2023mzc,Chakraborty:2023zdd},
a couple of comments are in order:
\begin{enumerate}
\item
For massless BTZ, $\rho_0=0$, in which case eq.~(\ref{h}) implies
\be\label{westring}
{\cal E}_{\rm string}=w{\cal E}_{w,string},
\ee
and the energies spectrum in~(\ref{recap}) takes the form:
\be\label{elambda}
E_w(\lambda)={w\over\lambda R}\left(-1+\sqrt{1+{2\lambda RE_w(0)\over w}+\left({\lambda RP\over w}\right)^2}\right),~\quad
P\equiv{n\over R}~,\quad \lambda\equiv{\alpha'\over R^2},
\ee
where $RE_w(0)=\lim_{\lambda\to 0}RE_w(\lambda)={\cal E}_{w,string}$.
The spectrum~(\ref{elambda}) has the pattern of a symmetric product of $T\bar T$ deformed $CFT_2$ with $c=6k$,~\footnote{The
central charge of the long string $CFT_2$,~\cite{Seiberg:1999xz}, denoted by ${\cal M}_{6k}^{(L)}$ in~\cite{Chakraborty:2019mdf},
can be determined in various ways; it will be re-determined in the comment on deformed global $AdS_3$ below.}
with the dimensionless single-trace $T\bar T$ coupling $\lambda$.~\footnote{Here, we restricted to $\lambda>0$;
the $\lambda<0$ case is obtained by taking $E\to iE$ and $R\to iR$, appropriately, above and below.}
This is the well known~\footnote{See e.g.~\cite{Giveon:2019fgr} and references therein.}
long strings spectrum of the superstring on deformed massless BTZ, denoted by ${\cal M}_3$
in~\cite{Giveon:2017nie,Giveon:2019fgr,Chakraborty:2019mdf,Chakraborty:2020swe}.
\item
For deformed massive BTZ, the excitation energy of a long string, $E_1(R)$,
plus its contribution to the energy of the black hole, ${E_{\rm deformed\,BTZ}(R)/p}$ in~(\ref{edefbtz}),
can be written (using~(\ref{mbtz})) as
\be\label{etotal}
E_{\rm total}(R)\equiv E_1(R)+{R\over\alpha'}\left(-1+\sqrt{1+{2\alpha'r_5M_{BTZ}\over R^2p}}\right).
\ee
This is the l.h.s. of~(\ref{recap}) (for $w=1$, and using~(\ref{mbtz})).
The total (dimensionless) energy corresponding to an excited long string in BTZ is
\be\label{etbtz}
{\cal E}_{\rm total}\equiv\lim_{{R\over\sqrt{\alpha'}}\to\infty}RE_{\rm total}(R)={\cal E}_{\rm string}+{r_5M_{BTZ}\over p}.
\ee
From the r.h.s. of~(\ref{recap}) (for $w=1$, and using~(\ref{mbtz})),
we see that $E_{\rm total}(R)$ is the energy in a $T\bar T$ deformed $CFT_2$ with $c=6k$
(with the dimensionless coupling defined in~(\ref{elambda}))
of a state in the $CFT_2$ with dimensionless energy ${\cal E}_{\rm total}$.
This is in harmony with the claim in~\cite{Giveon:2017nie,Chakraborty:2020swe,Chakraborty:2023mzc,Chakraborty:2023zdd}
that the black hole can be thought of as a state in a symmetric product of $p$ $T\bar T$-deformed $CFT_2$'s with $c=6k$,
where the black-hole energy is split equally among all $p$ factors (see below).
\item
Global $AdS_3$ amounts to $\rho_0^2=-r_5^2\equiv -\alpha' k$ in~(\ref{defbtz})--(\ref{withn}).
The long strings spectrum in the superstring on deformed global $AdS_3$ is thus given by eqs.~(\ref{recap})--(\ref{edefbtz})
with $\rho_0^2=-r_5^2$, and consequently,
\be\label{kp}
r_5M_{BTZ}\to -{c_p\over 12},\qquad c_p\equiv 6kp,
\ee
on the r.h.s. of~(\ref{mbtz}).~\footnote{See also \cite{CGH},
where we explore the spectrum of both discrete and continuum states in string theory on deformed global $AdS_3$ systematically.}
Note that $-c_p/12$ is the dimension less energy of the $SL(2,C)$-invariant ground state
in a $CFT_2$ with central charge $c_p$
(which is the central charge of the $CFT_2$ dual to string theory on $AdS_3$ in our case).
Now, following the discussion in the deformed BTZ case,
the total energy of an excited long string in deformed global $AdS_3$ is
\be\label{etads}
E_{\rm total}(R)=E_1(R)+{R\over\alpha'}\left(-1+\sqrt{1-{\alpha' k\over R^2}}\right).
\ee
This is the l.h.s. of~(\ref{recap}) (with $w=1$, and $\rho_0^2=-\alpha' k$).
In the global $AdS_3$ limit, we have
\be\label{egads}
{\cal E}_{\rm total}\equiv\lim_{{R\over\sqrt{\alpha'}}\to\infty}RE_{\rm total}(R)={\cal E}_{\rm string}-{c\over 12},\qquad c\equiv 6k.
\ee
From the r.h.s. of eq.~(\ref{recap}) (for $w=1$, and using~(\ref{mbtz}) with~(\ref{kp})),
we see that $E_{\rm total}(R)$ in eq.~(\ref{etads}) is the energy in a $T\bar T$ deformed $CFT_2$ with
central charge $c=c_p/p$,~(\ref{egads}),~(\ref{kp}),
of a state in the $CFT_2$ with dimensionless energy ${\cal E}_{\rm string}-{k\over 2}$
(with the dimensionless $T\bar T$ coupling defined in~(\ref{elambda})).
All in all, this is again in harmony with the symmetric product patterns of single-trace $T\bar T$ holography,
claimed in~\cite{Giveon:2017nie,Giveon:2019fgr,Chakraborty:2019mdf,Chakraborty:2020swe,Chakraborty:2023mzc,Chakraborty:2023zdd}
(see next section).
\end{enumerate}

\section{Summary of our results}

In conclusion, the picture that emerges is the following.
In~\cite{Giveon:2017nie,Chakraborty:2020swe,Chakraborty:2023mzc,Chakraborty:2023zdd},
it was assumed that there is an effective symmetric product picture of deformed BTZ black holes
(for instance, those formed by $p$ fundamental strings in the vicinity of $k$ NS5 branes,~(\ref{defbtz})--(\ref{withn})).
The claim is that a deformed BTZ black hole can be thought of as a state in a symmetric product
of $p$ $T\bar T$ deformed $CFT_2$’s of central charge $c=6k$, where  the black-hole (BH) energy,
\be\label{ebh}
E_{BH}\equiv E_{\rm deformed\,BTZ}
\ee
(see~(\ref{edefbtz})), is split equally among all $p$ factors.
A perturbative string carrying winding $w$ is an excitation in $w$ of the $p$ factors.
We shall discuss first the $w=1$ case, prior to inspecting string excitations carrying general $w\geq 1$.

A perturbative string carrying $w=1$ is an excitation of one of the factors.
Then, the energy of the state with a fundamental string excitation of energy
\be\label{eone}
E_1\equiv E_{\rm string}
\ee
(see above) is
\be\label{energy}
(p-1)E_{BH}/p+(E_{BH}/p+E_1),
\ee
where the last term in brackets is the contribution of the last of the $p$ factors -- the one with a string excitation.
Now, as we change the dimensionless $T\bar T$ coupling $\lambda$ in the block of the symmetric product,
each of the $p-1$ $E_{BH}/p$’s in the first term of~(\ref{energy})
evolves according to the formula~\footnote{Recall,~\cite{Smirnov:2016lqw,Cavaglia:2016oda},
that if the original $CFT_2$ is placed on a circle of radius $R$,
a state in the $CFT_2$ of energy $E$ and momentum $P$ gives rise in the $T\bar T$ deformed theory
to a state of energy $E(\lambda)$ (and the same momentum), with
$$
E(\lambda)={1\over\lambda R}\left(-1+\sqrt{1+2\lambda RE(0) +\left(\lambda RP\right)^2}\right),
$$
where $E(0)\equiv E$; $\lambda$ is related to the dimensionful $T\bar T$ coupling $t$, $\lambda\sim t/R^2$.}
\be\label{zamo}
E_{BH}(\lambda)/p={1\over\lambda R}\left(-1+\sqrt{1+2\lambda r_5M_{\rm BTZ}/p +\left(\lambda J_{\rm BTZ}\right)^2}\right),
\ee
with~\footnote{In section 2, we considered non-rotating deformed BTZ black holes, $J_{BTZ}=0$, for simplicity.}
\be\label{lrr}
\lambda\equiv{\alpha'\over R^2},
\ee
and the last factor in~(\ref{energy}), which we denote by
\be\label{defet}
E_{\rm total}(\lambda)\equiv E_{BH}/p+E_1
\ee
(see~(\ref{etotal})), evolves separately also according to~(\ref{zamo}),
\be\label{zamot}
E_{\rm total}(\lambda)={1\over\lambda R}\left(-1+\sqrt{1+2\lambda{\cal E}_{\rm total}+\lambda^2n^2}\right),
\ee
with
\be\label{mebtz}
{\cal E}_{\rm total}\equiv r_5M_{BTZ}/p+{\cal E}_{\rm string}
\ee
(see~(\ref{etbtz})).~\footnote{Note that~(\ref{zamo}) with $J_{BTZ}=n$ is a particular case of~(\ref{zamot}) with ${\cal E}_{\rm string}=0$.}

Finally, a perturbative string carrying winding $w$ is an excitation which involves $w$ of the $p$ factors
of the symmetric product.
Then, the energy of the state with a fundamental string excitation of energy
\be\label{eonew}
E_w\equiv E_{w\rm ,string}
\ee
(see above) is
\be\label{energyw}
(p-w)E_{BH}/p+(wE_{BH}/p+E_w),
\ee
where the last term in brackets is the contribution of the last of the $w$ factors -- those with a string excitation.
Now, as one changes the dimensionless $T\bar T$ coupling $\lambda$,~(\ref{lrr}), in the block of the symmetric product,
each of the $p-w$ $E_{BH}/p$’s in the first term of~(\ref{energyw}) evolves according to~(\ref{zamo})
and the last factor in~(\ref{energyw}), which we denote by
\be\label{defetw}
E_{w,\rm total}(\lambda)\equiv wE_{BH}(\lambda)/p+E_w(\lambda),
\ee
evolves separately, according to that in a $Z_w$ twisted sector of a symmetric orbifold,~\footnote{See e.g. the discussion around eqs.~(3.11)--(3.13) of~\cite{Giveon:2017myj}.}
\be\label{zamotw}
E_{w,\rm total}(\lambda)={w\over\lambda R}\left(-1+\sqrt{1+{2\lambda\over w}{\cal E}_{w,\rm total}+{\lambda^2n^2\over w^2}}\right),
\ee
with
\be\label{mebtzw}
{\cal E}_{w,\rm total}\equiv wr_5M_{BTZ}/p+{\cal E}_{w,\rm string}.
\ee
The result~(\ref{recap})--(\ref{edefbtz}), with~(\ref{h})$=$(\ref{jj}), has precisely this property.~\footnote{The
conclusion applies also to deformed global $AdS_3$,
with $E_{BH}\to E_{\rm deformed\,AdS}$, and correspondingly, $r_5M_{BTZ}\to -kp/2$ (see above).}
Hence, it is consistent with the picture that the long strings excitations deform together with the background
as a block of a symmetric product, in harmony with the single-trace $T \bar T$
holography conjecture of~\cite{Giveon:2017nie,Chakraborty:2020swe,Chakraborty:2023mzc,Chakraborty:2023zdd}.

\vspace{10mm}

\section*{Acknowledgments}
We thank David Kutasov for collaboration on many aspects of this paper.
The work of SC received funding under the Framework Program for Research and
“Horizon 2020” innovation under the Marie Skłodowska-Curie grant agreement no 945298. This work was supported in part by the FACCTS Program at the University of Chicago. The work of AG was supported in part by the ISF (grant number 256/22) and the BSF (grant number 2018068). The work of AH was supported in part by the U.S. Department of Energy, Office of Science, Office of High Energy Physics, under Award Number DE-SC0017647.

\vspace{10mm}

%\section*{Appendix}

\end{document}